\documentclass[12pt,a4paper]{iot-si}

\def\currentvolume{1}
\def\currentissue{1}
\def\currentyear{2009}
\def\currentmonth{April}
\def\ppages{0--0}

\usepackage{epsfig,subfigure}

\usepackage{array}
\usepackage{algorithm}
\usepackage{algorithmic}

\title[Computer-Aided Annotation for Video Tampering Dataset of Forensic Research]
      {Computer-Aided Annotation for Video Tampering Dataset of Forensic Research}
\author[Y. Yao]{}

\begin{document}

\maketitle

\centerline{Ye Yao}
 \medskip
{\footnotesize
\centerline{School of Cyberspace, Hangzhou Dianzi University}
\centerline{No.2 Street, Xiasha Higher Education Zone, Hangzhou, 310018, China}
\centerline{yyaoprivate@gmail.com} }
\medskip



\rule{15cm}{1pt}

\begin{abstract}
{\em The annotation of video tampering dataset is a boring task that takes a lot of manpower and financial resources. At present, there is no published literature which is capable to improve the annotation efficiency of forged videos. We presented a computer-aided annotation method for video tampering dataset in this paper. This annotation method can be utilized to label the frames of forged video sequences. By means of comparing the original video frames with the forged video frames, we can locate the position and the trajectory of the forged areas of the forged video frames. Then, we select several key points on the temporal domain according to the trajectory of the forged areas, and mark the forged area of the forged frames in the key point with a mouse. Finally, we use the linear prediction algorithm based on the coordinates of the key positions in the temporal domain to generate the annotation information of forged areas in other video frames which without manually labeled. If the bounding box generated by the computer-aided algorithm deviates from the actual location of the forged area, we can use the mouse to change the position of the bounding box during the preview period. This method combines the manual annotation with computer-aided annotation. It solves the problems of the inaccuracy of annotation by computer-aided as well as the low efficiency of annotation manually, and meet the needs of annotation for an enormous amount of forged videos in the research of video passive forensics.}

\end{abstract}

\rule{15cm}{1pt}

\section{Introduction}
Deep learning has been extensively used in the field of image and video processing, and has achieved a great success. At present, many researchers have begun to solve the problems of multimedia security~\cite{Rocha:2011}~\cite{Stamm:2013} such as steganalysis~\cite{Xu:2016} and digital image tampering detection~\cite{Qureshi:2015} in the field of multimedia security based on deep learning methods. Video tampering detection has drawn more and more attention in recent years. Some researchers have begun to use deep learning based methods to detect and locate the forged areas in the forged video sequences. However, detection and localization of video forged areas based on deep learning require a large number of labeled video samples for training. The video tampering dataset that is already available to researchers all over the world contains a small number of forged video sequences or does not provide the annotation information of forged areas. It can not meet the quantity requirement of the training data that required for deep learning.

\begin{figure}[htbp]
\begin{center}
\subfigure[GOP structure of original video sequence (GOP size=25)]{
\includegraphics[width=14cm]{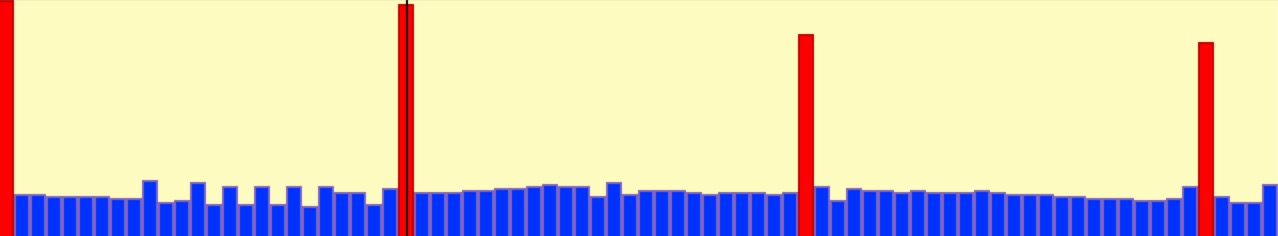}
}
\subfigure[GOP structure of forged video sequence (GOP size=33)]{
\includegraphics[width=14cm]{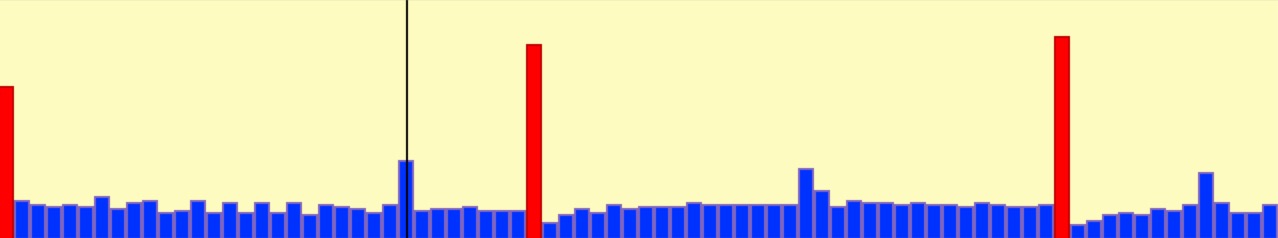}
}
\caption{Different GOP structure and frame size between original and forged}
\label{fig:gop}
\end{center}
\end{figure}

The computer-aided annotation~\cite{Real:2017}~\cite{Cuevas:2016} for video tampering dataset cannot merely adopt the algorithms in the fields of computer vision and image processing. Firstly, the forged areas may be larger than the areas where the moving object is removed or inserted. Therefore, by means of comparing the difference between the moving objects in the original video frames and in the forged video frames, we can not get the location information of forged areas on the temporal domain and on the spatial domain accurately. Secondly, after the moving object in the video sequence has been forged, the video frames have to be compressed again, which is named as double compression. This will bring noise and may also change the GOP structure of the video sequence. As shown in Fig.~\ref{fig:gop}, the GOP size in the original video sequence is 25, but it changes to 33 in the forged video sequence after double compression. Because of the noise introduced by the compression, subtracting the original video frames from the forged video frames cannot get the location information of the forged areas. Therefore, an algorithm has to be designed for the annotation of video tampering dataset. This algorithm should be able to eliminate the noise brought by video compression and output location information of forged areas accurately.

A computer-aided annotation method for video tampering dataset is presented in this paper. This annotation method can be utilized to label the frames of forged video sequences. By means of comparing the original video frames with the forged video frames, we can locate the position and the trajectory of the forged areas in the forged video frames. Then, we select several key points on the temporal domain according to the trajectory of the forged areas, and mark the forged area of the forged frames in the key point with a mouse. Finally, we use the linear prediction algorithm based on the coordinates of the key positions in the temporal domain to generate the annotation information of forged areas in other video frames which without manually labeled. If the bounding box generated by the computer-aided algorithm deviates from the actual location of the forged area, we can use the mouse to change the position of the bounding box during the preview period. This method combines the manual annotation with computer-aided annotation. It solves the problems of the inaccuracy of annotation by computer-aided as well as the low efficiency of annotation manually and meets the need of annotation for an enormous amount of forged videos in the research of video passive forensics. 

\section{Related Work}
\subsection{Datasets}
There are several video tampering dataset for video forensics research. As far as we know, the first video tampering dataset is SULFA(Surrey University Library for Forensic Analysis)~\cite{Qadir:2012}, which was designed and built by the forensics research group at Surrey University. SULFA contains several original and several forged video sequences, which is freely available through the website of research group\cite{sulfa:2012}. There are five pairs of video sequences with copy-move forgery. Each forged video includes information of the forged area namely as annotation in this paper.

REWIND tampering dataset\cite{rewinddb:2013} was built by Bestagini~\emph{et al.}. This dataset is composed of 10 original video sequences and 10 forged ones. Each sequence has a resolution of 320x240 pixels, and a frame-rate of 30 fps. Some of the original sequences come from the SULFA tampering dataset. REWIND also contains the differences between the frames of the original sequences and the forged sequences, which serves as the annotation information.

Al-Sanjary~\emph{et al.} developed a video tampering dataset for forensic research\cite{Al-Sanjary:2016}. It includes 10 video sequences with copy-move tampering. Some information such as the video length, the total number of frames and the ID number of forged frames are provided in the ground truth, but there is no information about the tampering areas in the temporal domain.

Although SULFA and REWIND include the annotation information, the quantity of forged video sequences in these datasets is too low. It can not meet the requirement of deep learning based forensic method. SYSU-OBJFORG is the largest object-based forged video dataset according to the report in~\cite{Chen:2016}. It consists of 100 pristine video sequences and 100 forged video sequences. All video sequences are of 11 seconds, 1280$\times$720 H.264/MPEG-4 encoded sequences with a bitrate of 3 Mbit/s and a frame rate of 25. But there is no annotation information provided with the SYSU-OBJFORG dataset. In order to extract positive samples and negative samples from SYSU-OBJFORG dataset for the training of deep learning, an annotation algorithm should be presented to help to mark the forged areas in the forged video frames.

\subsection{Software and tool}
With the wide availability of user-friendly media editing software such as Premiere, After Effects and Final Cut, it becomes much easier to insert video objects to video sequences or remove video objects from video sequences. However, these pieces of software cannot export coordinate information of the forged area in each video frame.

Ardizzone~\emph{et al.} present a tool for forensics researcher to create forged videos\cite{Ardizzone:2015}. The author also created a forged video dataset from 6 video sequences. This dataset is available at\cite{unipa:2015} with the information of bounding boxes for the modified objects. This tool can be utilized to clone video objects from a video frame to another video frame and generate a forged video, but this tool can not be used to calculate the difference between the original video sequence and the forged video sequence. That is to say, we can not use this tool to create an annotation for a existed tampering dataset such as SYSU-OBJFORG.

As a result, we present a computer-aided annotation method for video tampering dataset in this paper. This annotation method can be utilized to label the frames of a forged video sequence. It is a solution to the annotation of video tampering dataset which can significantly improve the accuracy and efficiency.

\section {Method}
\subsection{Select the key point}
\begin{figure}[htbp]
\begin{center}
\includegraphics[width=12cm]{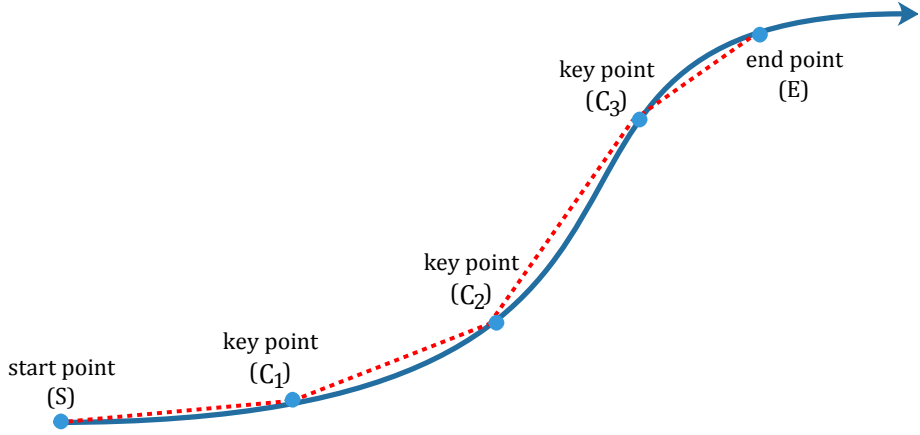}
\caption{The top view for the actual trajectory of the forged areas (Solid blue line) and the manually annotation point (key point).}
\label{fig:trajectory}
\end{center}
\end{figure}

The first step in our method is to select the key point in the trajectory of the forged object. This method requires us to choose the key points manually. We are required to preview the original video sequence and the forged video sequence to obtain the rough trajectory based on our estimation. By means of comparing the difference between the original video frame and the forged video frame with our naked eyes, we can find out the forged area in each forged video frame. If we take each forged area as a point on the temporal domain, all of the forged areas in the forged video sequence compose a trajectory of the forged object.

As shown in Fig.~\ref{fig:trajectory}, the solid blue line is the actual trajectory of a forged object. Between the start point and the end point of the trajectory, we can select some points as key points. At each key point, there is a forged video frame. We need to mark out the forged areas on the forged video frame with a mouse. In order to let the prediction results more closely to the actual trajectory, we need to select out more key points. On the other hand, the more key points we select, the more time we need to spend on. So we have to be a trade-off between prediction accuracy and manpower cost. We have given an example of the selection of the key point in Fig.~\ref{fig:trajectory}. As shown in Fig.~\ref{fig:trajectory}, we select a point as the key point when the forged object changes its trajectory in an unexpected speed or to an uncertain direction. It can greatly reduce the manpower cost for annotation as well as keep the prediction accuracy in a good condition.

\begin{figure}[htbp]
\begin{center}
\includegraphics[width=12cm]{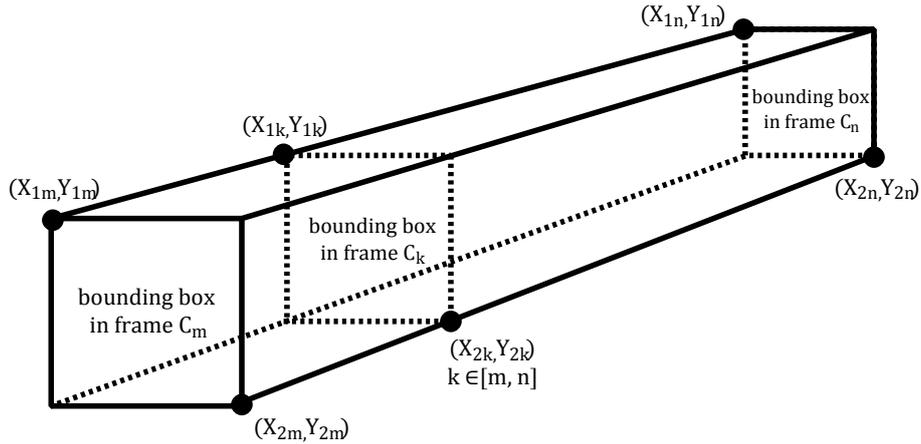}
\caption{Predict bounding-box in frame $k$ based on bounding boxes in frame $m$ and frame $n$.}
\label{fig:predict}
\end{center}
\end{figure}

\subsection{Predict other points}
After we finished the selection of the key point, we need to predict the tampering area in other forged frames between any two of the key points. As shown in Algorithm~\ref{alg1} and Fig.\ref{fig:trajectory}, we divide the trajectory into several segments. For each segment, it is starting with a key point and ending with a consecutive key point. We apply our prediction algorithm to each segment one by one.

\begin{algorithm}
  \caption{Predict other points between start point $S$ and end point $E$}
  \label{alg1}
  \begin{algorithmic}
  \REQUIRE start point $S$, key points $\{C_1, C_2, ..., C_n\}$, end point $E$
  \ENSURE prediction result $C_k, k \in [S, E]$
  \STATE start key point $m \gets S$
  \FOR{$idx$ in $\{C_1, C_2, ..., C_n, E\}$}
  \STATE end key point $n \gets idx$
  \STATE predict all of the other point $C_k$ between $[m, n]$
  \STATE start key point $m \gets idx$
  \ENDFOR
  \end{algorithmic}
\end{algorithm}

Between the key point $m$ and $n$, we use Formula (\ref{formula1}) to get the coordinate of the bounding box in the forged frame $C_k$. Note that according to Fig.\ref{fig:predict}, the frame $C_k$ is between the frame $C_m$ and the frame $C_n$. Therefore, the prediction resulting $(X_{1k}, Y_{1k})$ and $(X_{2k}, Y_{2k})$ can be regarded as the coordinate of the bounding box in frame $C_k$, which represents the forged area in the forged video frame $C_k$.

\begin{equation}\label{formula1}
\begin{split}
&X_{1k} = X_{1m} + (X_{1n}-X_{1m})\times(k-m)/(n-m)\\
&Y_{1k} = Y_{1m} + (Y_{1n}-Y_{1m})\times(k-m)/(n-m)\\
&X_{2k} = X_{2m} + (X_{2n}-X_{2m})\times(k-m)/(n-m)\\
&Y_{2k} = Y_{2m} + (Y_{2n}-Y_{2m})\times(k-m)/(n-m)
\end{split}
\end{equation}
where $m, n, k$ are the frame number in the forged video sequence, $(X_{1m}, Y_{1m})$ and $(X_{2m}, Y_{2m})$ are the left top coordinate and the right bottom coordinate of the bounding box in frame $C_m$, $(X_{1n}, Y_{1n})$ and $(X_{2n}, Y_{2n})$ are the left top coordinate and the right bottom coordinate of the bounding box in frame $C_n$.

\begin{figure}[htbp]
\begin{center}
\includegraphics[width=15cm]{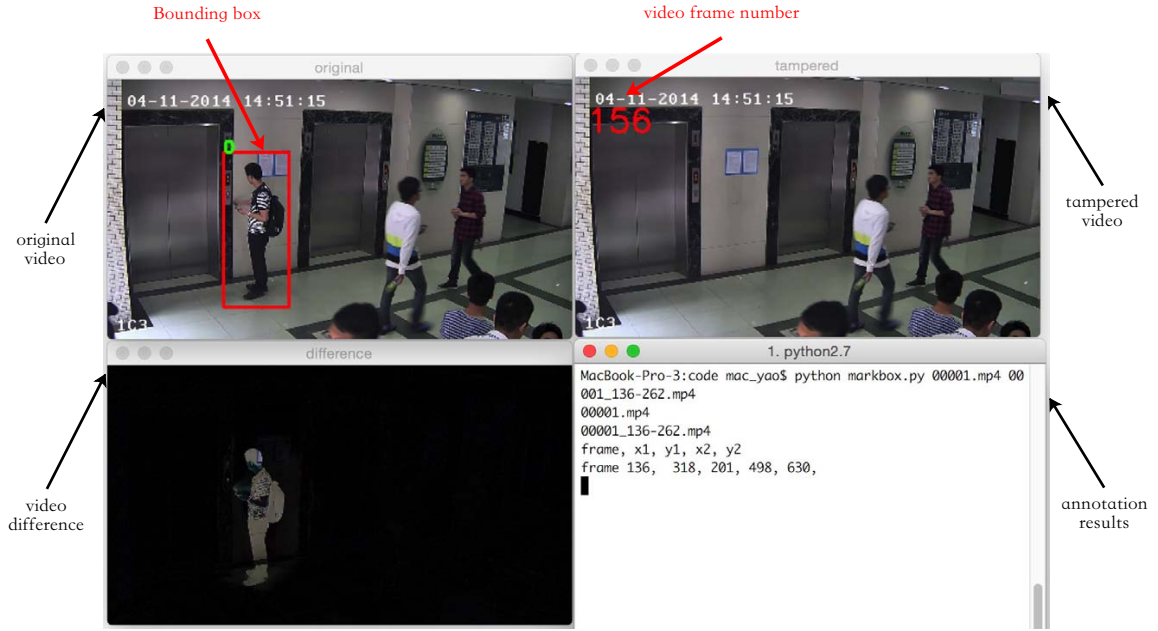}
\caption{Graphical user interface.}
\label{fig:ui}
\end{center}
\end{figure}

\section {Results}
\subsection{Tool}
The source code corresponding to the annotation method has been freely available through github~\cite{sourcecode:2018}. We implement the algorithm with python and OpenCV. The graphical user interface is shown in Fig.\ref{fig:ui}. There are four windows. Three of them display the original video sequence, the forged video sequence and the frame difference between the original and the forged respectively. The last one displays the information of the bounding box outputting by the annotation tool.

To let users more easily to find out the forged area, we show the frame difference between the original and the forged on the third window. The difference is the result of subtraction between the two video frames. Based on the images shown in these windows, the user can mark out all the forged area in the forged video sequence with a mouse.

We also add some codes to support the modification of the bounding box. The user can delete one bounding box by clicking on the bounding box through the right button of the mouse. After removing this bounding box, a new bounding box can be drawn on the correct area through the left button of the mouse. All of the operations are responded to the first window, namely, are processed through the window of the original video sequence. 

There are several short keys for users to have a faster and smoother experience. For example, users can save the bounding box to the annotation list by pressing the ``S'' key. Users also can go back to the last video frame by pressing the ``R'' key, or display the next video frame by pressing any other keys.

\renewcommand\arraystretch{1.2}
\begin{table}[htbp]
	\caption{Statistics of the annotation result for the SYSU-OBJFORG}
	\label{table1}
	\begin{center}
    \begin{tabular}{|p{6.5cm}|p{2.5cm}<{\centering}|p{2.5cm}<{\centering}|}
    \hline %
    \quad  \quad  \quad  \quad  \quad  \quad Item & Number & Ratio \\
    \hline \hline
    Tampered video sequences & 100 & \\
    \cline{1-2}
    Original video sequences & 100 & \\
    \hline \hline
    Total tampered frames ($F_1$)& 11074 & $F_1/F_2=$ \\
    \cline{1-2}
    Total original frames ($F_2$)& 47368 & 23.38\% \\
    \hline \hline
    Bounding box marked ($B_1$) & 586 & $B_1/B_2=$ \\
    \cline{1-2}
    Bounding box total ($B_2$) & 11837 & 4.95\% \\
    \hline \hline
    Total time of marking (in Hours) & 3.5 hours &  \\
    \hline %
    \end{tabular}
    \end{center}
\end{table}

\subsection{Evaluation}
We test our algorithm and tool on the SYSU-OBJFORG dataset, which is the biggest video tampering dataset till now. The annotation result is also available at github~\cite{sourcecode:2018} to researchers who want to train their tampering detection models. The statistics of the annotation results of the SYSU-OBJFORG dataset are shown in Table \ref{table1}. It only takes 3.5 hours to mark all of the 100 pairs video sequences. There is no more than 5\% bounding box that needs to be marked manually. That is to say, all of the other bounding boxes are generated by the computer-aided annotation algorithm. This algorithm is important for the researchers to annotate the forged video sequence quickly.

\section {Conclusions}
In this paper, a computer-aided annotation method for video tampering dataset has been presented. This method is developed for the purpose of marking the forged area of the forged video sequence. With the aid of computer algorithms, we can increase the efficiency of annotation greatly. For the future work, we will continue to develop our method to support pixel-level marking. It will have the capability to annotate the forged area more precisely. The availability of such tools and algorithms will be extremely useful for forensics researchers. It can provide them with better training samples and improve the detection results of the algorithms based on deep learning.

\section*{Acknowledgment}
This work is partially supported by the Humanities and Social Sciences Foundation of Ministry of Education of China under Grant 17YJC870021, partially supported by the Public Technology Application Research Project of Zhejiang Province under Grant 2017C33146.
The authors also gratefully acknowledge the helpful comments and suggestions of the reviewers, which have improved the presentation.


\end{document}